\documentclass[11pt]{article}

\usepackage[text={17.1cm,24.6cm},centering]{geometry} 
\usepackage[utf8]{inputenc}
\usepackage{authblk}
\usepackage{blkarray}
\usepackage[column=S]{cellspace} 
\usepackage{amsfonts,amsmath,amssymb,amsthm}
\usepackage{latexsym,mathrsfs,mathtools,bm}
\usepackage{graphicx,subcaption,epsfig,caption,float,xcolor}
\usepackage{enumitem}
\usepackage{chngcntr}
\usepackage[column=S]{cellspace} 
\usepackage{tabularx}            
\usepackage[colorlinks=true,linkcolor=blueG, citecolor=redG, urlcolor=magentaG, bookmarks]{hyperref}

\newcommand{\nc}{\newcommand}
\nc{\ir}{\mathrm{i}}
\nc{\dd}{\mathrm{d}} 
\nc{\eE}{\mathrm{e}}
\nc{\Jp}{t^+}
\nc{\Jm}{t^-}
\nc{\Jpm}{t^\pm}
\def\rr{{\boldsymbol{\rho}}}
\def\orr{\overline{{\boldsymbol{\rho}}}}
\def\ox{{\overline{x}}}

\definecolor{blueG}{RGB}{51, 102, 204}
\definecolor{magentaG}{RGB}{214.2, 40.8, 132.6}
\definecolor{redG}{RGB}{229.5, 51., 102.}

\usepackage{tikz}
\usetikzlibrary{calc}
\usepackage{cite}
\usepackage{bookmark}

    \def\cN{{\cal N}}

\numberwithin{equation}{section}
\theoremstyle{plain}
\newtheorem{thm}{Theorem}[section]

\newtheorem{prop}[thm]{Proposition}

\newtheorem{rmk}[thm]{Remark}

\title{\bf Inhomogeneous SSH  models and \\ the doubling of orthogonal polynomials}
\renewcommand*{\Affilfont}{\normalsize\small}

\author[1]{Nicolas Cramp\'e\,}
\author[2]{Quentin Labriet\,}
\author[2]{Lucia Morey\,}
\author[3]{Gilles Parez\,}
\author[2,5]{Luc Vinet\,\vspace{.5em}}
\affil[1]{\textit{CNRS -- Universit\'e de Montr\'eal CRM - CNRS, Montr\'eal (Qu\'ebec), H3C 3J7, Canada\newline}}
\affil[2]{\textit{Centre de Recherches Math\'ematiques, Universit\'e de Montr\'eal, P.O. Box 6128,
\newline\vspace{.9em}
 Centre-ville Station, Montr\'eal (Qu\'ebec), H3C 3J7, Canada}}
\affil[3]{\textit{Laboratoire d'Annecy de Physique Th\'eorique (LAPTh), CNRS,\newline\vspace{.9em} Universit\'e Savoie Mont Blanc, 
  74940~Annecy, France}}
\affil[5]{\textit{IVADO, Montr\'eal (Qu\'ebec), H2S 3H1, Canada \newline\vspace{.9em}}}

{
 \makeatletter
 \renewcommand\AB@affilsepx{: \protect\Affilfont}
 \makeatother
 \makeatletter
 \renewcommand\AB@affilsepx{, \protect\Affilfont}
 \makeatother
}
\date{\today}

\begin{document}
\maketitle

\begin{abstract}
We analyze Su--Schrieffer--Heeger (SSH) models using the doubling method for orthogonal polynomial sequences. This approach yields the analytical spectrum and exact eigenstates of the models. We demonstrate that the standard SSH model is associated with the doubling of Chebyshev polynomials. Extending this technique to the doubling of other finite sequences enables the construction of Hamiltonians for inhomogeneous SSH models which are exactly solvable. We detail the specific cases associated with Krawtchouk and $q$-Racah polynomials. This work highlights the utility of polynomial-doubling techniques in obtaining exact solutions for physical models.
\end{abstract}
\newpage
\tableofcontents

\section{Introduction}

Exactly-solvable models of quantum-many body systems play a significant role in our understanding of emergent phenomena, such as long-range order \cite{lieb1961two} and quantum phase transitions \cite{OAFF02,ON02,VLRK03, CC04}, entanglement measures \cite{peschel2003calculation,its2005entanglement,peschel2009reduced}, relaxation dynamics out of equilibrium \cite{cc-05,fc-08,parez2022analytical}, or the impact of spatial inhomogeneities on physical properties \cite{eisler2009entanglement,DSVC17}. 

The Su--Schrieffer--Heeger (SSH) model is a prominent example of such models, and was originally introduced to describe polyacetylene~\cite{SSH}, where carbon atoms form a dimerized chain with alternating strong and weak bonds. Owing to this alternating bonding pattern, the model hosts symmetry-protected topological excitations and represents the simplest quantum many-body system exhibiting such features \cite{chiu2016classification}. The SSH model is exactly solvable with free-fermion techniques, but nonetheless describes experimentally realized topological phases \cite{meier2016observation,leder2016real,huda2020tuneable}, making it a paradigmatic framework for studying topological phase transitions in condensed matter systems. 

In this paper, we use the mathematical properties of orthogonal polynomials and their doubling to revisit the solution of the standard SSH model and construct different solvable inhomogeneous generalizations. A strong relationship exists between orthogonal polynomials of the Askey scheme and free-fermion models. This connection has already yielded exact results for inhomogeneous XX or XY spin chains \cite{crampe2019free,BernardXY}, and free-fermion models on various graphs \cite{BernardHamming,BernardJohnson,ParezMulti,bernard2023absence} or higher-dimensional lattices \cite{bernard2022entanglement}. Moreover, it allows for the analytical investigation of physical properties, such as perfect state transfer or entanglement measures \cite{VinetPST,bernard2023entanglement,blanchet24Neg,bernard2024entanglement,bernard2025entanglement}.

The doubling procedure for orthogonal polynomials consists in combining two families of orthogonal polynomials in an appropriate way to produce a new one \cite{MARCELLAN1997,Oste2015Doubling}. This approach allows one to construct and diagonalize a symmetric tridiagonal matrix whose eigenvectors are expressible in terms of the original polynomial families. The resulting tridiagonal matrix has particularly nice features: the elements on the sub- and super-diagonals alternate between two sequences, $\Jp_n$ and $\Jm_n$, describing the alternating strength of the bonds and their inhomogeneity along the chain. We show that this construction provides a new perspective on the standard solution of the SSH model and enables the definition of a broader class of exactly solvable inhomogeneous SSH-like models.

While the inhomogeneous couplings generated by orthogonal-polynomial constructions do not aim to describe a specific crystalline material, they naturally correspond to engineered hopping profiles. Such spatially modulated SSH-type models are experimentally relevant in platforms where couplings can be controlled with high precision, including cold atoms in optical lattices, photonic or acoustic lattices, and other synthetic quantum systems \cite{meier2016observation,leder2016real,huda2020tuneable}. In this sense, these models provide analytically tractable realizations of inhomogeneous topological chains rather than microscopic, material-specific descriptions.

\paragraph{Main results.}The Hamiltonian of the inhomogeneous models studied in this paper has the form
\begin{equation}\label{eq.HSSHinh}
    \mathcal{H} = \sum_{n=1}^N \big(t_{n-1}^+ c_{2n-2}^\dagger c_{2n-1}+t_{n-1}^- c_{2n-1}^\dagger c_{2n} +\textrm{h.c.}\big),
\end{equation}
where $t^\pm_n$ are parameters characterizing the strength of the interaction between neighboring sites, and $c_n^{(\dagger)}$ are the fermionic operators satisfying the canonical anticommutation relations
\begin{equation}\label{eq:antic}
    \{c_m^\dagger, c_n\} = \delta_{m,n},\  \{c_m, c_n\} = 0 =  \{c_m^\dagger, c_n^\dagger\}\;,\quad (m,n=0,1,\dots, N).
\end{equation}
We represent a schematic inhomogeneous SSH model in Fig.~\ref{fig:graph2}.
\begin{figure}
\begin{center}
\begin{tikzpicture}[scale=0.8]
\draw[-] (0,0)--(7,0);\draw[dashed] (7,0)--(11,0);
\draw[] (11,0)--(16,0);
\foreach \s in {0,2,...,6,12,14,16} {
\draw [fill] (\s,0) circle (0.08);  
};
\node[above] at (0,0) {0};
\node[below] at (1,0) {$t^+_0$};
\node[above] at (2,0) {1};
\node[below] at (3,0) {$t^-_0$};
\node[above] at (4,0) {2};
\node[below] at (5,0) {$t^+_1$};
\node[above] at (6,0) {3};
\node[above] at (11.7,0) {$2N-2$};
\node[below] at (13,0) {$t^+_{N-1}$};
\node[above] at (14,0) {$2N-1$};
\node[below] at (15,0) {$t^-_{N-1}$};
\node[above] at (16,0) {$2N$};
\end{tikzpicture}
\end{center}
\caption{
Representation of an inhomogeneous SSH model.
\label{fig:graph2}  }
\end{figure}
The doubling procedure allows us to obtain the coupling constants $t^\pm_n$ such that the spectrum and the eigenvectors can be computed exactly. The different models obtained, with the names of the polynomials used in the doubling procedure, are summarized in the Table \ref{Tbl:summary} given below.

\paragraph{Outline of the paper.}

The paper is organized as follows: in Sec.~\ref{sec:SSH-chebyshev} we
show that the eigenvalues and eigenvectors of the standard SSH model can be related to the well-known Chebyshev polynomials and to their doubling.
Section \ref{sec:SSS-doubling} extends the doubling procedure to any finite sequence of orthogonal polynomials, leading to new models that generalize the SSH model while remaining exactly solvable. Finally, in Secs.~\ref{sec:Krawtchouk} and \ref{SSH-qRacah}, we illustrate the general construction using two specific families of discrete polynomials from the ($q$-)Askey scheme: the Krawtchouk and $q$-Racah polynomials.

\paragraph{Notations.} We recall here the definitions of the generalized ($q$-)hypergeometric functions used in this paper. The  generalized hypergeometric functions \cite{GasperRahman2004,KoekoekLeskyetal2010} are defined as
\begin{align}
{}_{r+1}F_r\left({{-n,\;a_1,\; a_2,\; \cdots,\; a_{r}  }\atop
{b_1,\; b_2,\; \cdots,\; b_{r} }}\;\Bigg\vert \; z\right)=\sum_{k=0}^{n}
\frac{(-n,a_1,a_2,\cdots,a_r)_k}{k!(b_1,b_2,\cdots,b_r)_k}z^k\,,
\end{align}
for $r,n$ non-negative integers,
and where the Pochhammer symbols are
\begin{align}
(b_1,b_2,\cdots,b_r)_k=(b_1)_k(b_2)_k\cdots (b_r)_k\,,\qquad (b)_k=b(b+1)\cdots (b+k-1)\,.
\end{align}
The $q$-hypergeometric functions \cite{GasperRahman2004,KoekoekLeskyetal2010} are defined by
\begin{align}
\label{eq:q-hypergeom-func}
{}_{r+1}\phi_r\left({{q^{-n},\;a_1,\; a_2,\; \cdots,\; a_{r}  }\atop
{b_1,\; b_2,\; \cdots,\; b_{r} }}\;\Bigg\vert \; q;z\right)=\sum_{k=0}^{n}
\frac{(q^{-n},a_1,a_2,\cdots,a_r;q)_k}{(b_1,b_2,\cdots,b_r,q;q)_k}z^k\,,
\end{align}
for $r,n$ non-negative integers,
with the $q$-Pochhammer symbols given by
\begin{align}
(b_1,b_2,\cdots,b_r;q)_k=(b_1;q)_k(b_2;q)_k\cdots (b_r;q)_k\,,\quad (b;q)_k=(1-b)(1-qb)\cdots (1-q^{k-1}b)\,.
\end{align}

\section{SSH model and doubling of Chebyshev polynomials \label{sec:SSH-chebyshev}}

This section is devoted to study the standard SSH model, where the coupling constant are constant along the chain: $t^+_n=t^+$ and $t^-_n=t^-$.
We consider a chain with an odd number of sites, labeled from $0$ to $2N$ and endowed with open boundary conditions. The Hamiltonian $\mathcal{H}$ for the homogeneous SSH model reads \cite{SSH}
\begin{equation}\label{eq.HSSH}
\begin{split}
    \mathcal{H} =& \sum_{n=1}^N \big(t^+ c_{2n-2}^\dagger c_{2n-1}+t^- c_{2n-1}^\dagger c_{2n} +\textrm{h.c.}\big)\\
    =& \sum_{m,n=1}^{2N+1}H_{m,n} c_{m-1}^\dagger c_{n-1},
    \end{split}
\end{equation}
where $c_n^{(\dagger)}$ are the fermionic operators satisfying the canonical anticommutation relations given in Eq.~\eqref{eq:antic}. 
The dimerized interactions $t^\pm$ read
\begin{equation}
    t^\pm = \frac{1\pm \delta}{2}
\end{equation}
with $-1 \leqslant \delta \leqslant 1$. 
We represent the chain described by this Hamiltonian in Figure \ref{fig:graph}, where the alternating bounds are easily visualized.

\begin{figure}
\begin{center}
\begin{tikzpicture}[scale=0.8]
\draw[-] (0,0)--(7,0);\draw[dashed] (7,0)--(11,0);
\draw[] (11,0)--(16,0);
\foreach \s in {0,2,...,6,12,14,16} {
\draw [fill] (\s,0) circle (0.08);  
};
\node[above] at (0,0) {0};
\node[below] at (1,0) {$t^+$};
\node[above] at (2,0) {1};
\node[below] at (3,0) {$t^-$};
\node[above] at (4,0) {2};
\node[below] at (5,0) {$t^+$};
\node[above] at (6,0) {3};
\node[above] at (11.7,0) {$2N-2$};
\node[below] at (13,0) {$t^+$};
\node[above] at (14,0) {$2N-1$};
\node[below] at (15,0) {$t^-$};
\node[above] at (16,0) {$2N$};
\end{tikzpicture}
\end{center}
\caption{
Representation of the usual SSH model with open boundary conditions.
\label{fig:graph}  }
\end{figure}

Because the full Hamiltonian $\mathcal{H}$ is quadratic in terms of the fermionic operators, it is well-known that it suffices to diagonalize a $(2N+1) \times (2N+1)$ matrix $H$ to obtain the eigenvalues and eigenvectors of $\mathcal{H}$. This matrix~$H$ is the single-particle Hamiltonian, and it reads
\begin{equation}
\label{eq:matrixH-ssh}
 H=\left( \begin{array}{c c c c c c c}
    0& t^+ & \\
    t^+ & 0 & t^-\\
    0 &t^- & 0 & t^+ \\
    &&& \ddots\\
    &&&t^+ & 0 & t^- \\
    &&&&t^- & 0
    \end{array} \right).
\end{equation}

\subsection{Diagonalization by Chebyshev polynomials \label{sec:diaCh}}

The diagonalization of this homogeneous SSH model is standard, see e.g. \cite{shin1997formula, sirker2014boundary,asboth2016short}. However,
to obtain exactly solvable inhomogeneous SSH model, we start by showing that the eigenvalues and eigenvectors of the homogeneous SSH model can be obtained using the doubling property of Chebyshev polynomials~\cite{MARCELLAN1997}. This offers a new mathematical framework for the standard diagonalization of the homogeneous model, and it extends naturally to inhomogeneous cases.

Let us recall some basic properties of the Chebyshev polynomials of the second kind, denoted as $U_n$ ($n=0,1,2,\dots$). They can be expressed in terms of hypergeometric functions as 
\begin{align}
    U_n(x)=(n+1)\ {}_2F_1 \left({{ -n,\;n+2}\atop
{\frac{3}{2}}}\;\Bigg\vert  \frac{1-x}{2}\right)\,.
\end{align}
Equivalently, using the substitution $x=\cos\theta$, they can be rewritten in trigonometric form,
\begin{equation}
U_n(\cos\theta)=\frac{\sin((n+1)\theta)}{\sin\theta}\,.
\end{equation}
With the trigonometric form, it is immediate to obtain the roots $x_k$ of $U_n$,
\begin{equation}
\label{eq:roots-Un}
x_k=\cos\left(\frac{k\pi}{n+1}\right), \quad (k=1,\ldots,n).
\end{equation}
Another well-known expression of $U_n$ is
\begin{align}\label{eq:Uexp}
    U_n(x)=\frac{\left(x+\sqrt{x^2-1}\right)^{n+1}-\left(x-\sqrt{x^2-1}\right)^{n+1}}{2\sqrt{x^2-1}}\,.
\end{align}
The Chebyshev polynomials of the second kind satisfy the three-term recurrence relation
\begin{equation}\label{eq:recuU}
     U_0(x)=1\,,\quad  2xU_{n}(x)=U_{n+1}(x)+U_{n-1}(x)\,, \quad (n=0,1,2\dots)\,,
\end{equation}
with the convention $U_{-1}=0$.
We define a new sequence of polynomials $(Q_n)_n$ as 
\begin{align}\label{eq:Qeven}
 &  Q_{2n}(x)= \frac{1+\delta}{2} U_{n}(\pi_x)+\frac{1-\delta}{2} U_{n-1}(\pi_x)\,,\\
 \label{eq:Qodd}
   & Q_{2n+1}(x)=xU_{n}(\pi_x)\,, 
\end{align}
where $\pi_x$ is given as
\begin{align}\label{eq:pi1}
  \pi_x=\frac{2}{1-\delta^2}x^2-\frac{1+\delta^2}{1-\delta^2} \,.
\end{align}
Note that $Q_n$ is a polynomial of degree $n$ in $x$.
From the definition of $Q_n$, it follows that
\begin{align}
\label{eq:chebyshev-xQ2n}
   x Q_{2n}(x)= \frac{1+\delta}{2} Q_{2n+1}(x)+\frac{1-\delta}{2} Q_{2n-1}(x)\,,
\end{align}
and 
\begin{align}
\label{eq:chebyshev-xQ2n1}
   x Q_{2n+1}(x)&=x^2U_{n}(\pi_x)=\left(\frac{1-\delta^2}{2}\pi_x+\frac{1+\delta^2}{2}\right) U_{n}(\pi_x) \nonumber\\
   &=\frac{1-\delta^2}{4}(U_{n+1}(\pi_x)+U_{n-1}(\pi_x))+\frac{1+\delta^2}{2} U_{n}(\pi_x)\nonumber\\
   &=\frac{1-\delta}{2} Q_{2n+2}(x)+\frac{1+\delta}{2} Q_{2n}(x)\,.
\end{align}
To obtain the second line in the above relation, we have used the recurrence relation \eqref{eq:recuU} satisfied by the Chebyshev polynomials $U_n(\pi(x))$. 
From Eqs.~\eqref{eq:chebyshev-xQ2n} and \eqref{eq:chebyshev-xQ2n1}, we verify that the vector $\bold{Q}(x)$ given by
\begin{align}
    \bold{Q}(x)=\big(Q_0(x),Q_1(x),\dots, Q_{2N}(x)\big)^t\,,
\end{align}
is an eigenvector of $H$ \eqref{eq:matrixH-ssh} corresponding to the eigenvalue $x$, where $x$ is a root of the characteristic polynomial:
\begin{align}
    Q_{2N+1}(x)=xU_{N}(\pi_x)=0\,.
\end{align}
Therefore, knowing the roots of $U_{N}$ in Eq.~\eqref{eq:roots-Un}, the eigenvalues are given by 
\begin{align}
    0, \ x_k^\pm=\pm \sqrt{\frac{1-\delta^2}{2}\cos\left(\frac{k\pi}{N+1}\right)+\frac{1+\delta^2}{2}}\,,\quad (k=1,2,\dots,N)\,.
\end{align}

The eigenvector associated to the eigenvalue $x=0$ is known as a \textit{zero mode}, which is a fundamental feature of topological insulators \cite{asboth2016short,chiu2016classification}. In the standard SSH model with odd number of sites, the zero mode is localized at the right (resp. left) boundary of the chain for $\delta>0$ (resp. $\delta<0$). These two situations correspond to distinct topological phases of the model. For $\delta=0$, the model is instead critical, or gapless.  

It is important to emphasize that the parity of the number of sites plays a crucial role in obtaining analytical results. Indeed, when the number of sites is even, equal to $2N$, the corresponding equation to determine the spectrum would be $Q_{2N}(x)=0.$ However the roots of this polynomial are not known explicitly in general.

\begin{rmk}\label{rem:mar}
The construction of the sequence $(Q_n)_n$ can be seen as the doubling procedure introduced in \cite{MARCELLAN1997}.
Indeed, let us introduce the sequence of polynomials $(V_n)_n$ defined by 
\begin{align}\label{eq:P1}
    V_n(x)=\frac{1+\delta}{2}U_n(x)+\frac{1-\delta}{2}U_{n-1}(x)\,.
\end{align}
One can show that the sequence $(Q_n)_n$ is constructed from $(V_n)_n$ as
  \begin{align}
   & Q_{2n}(x)=V_n(\pi_x)\,,\\
   &  Q_{2n+1}(x)=\frac{x}{(1+\delta)(\pi_x-c)}\left(V_{n+1}(\pi_x)-\frac{V_{n+1}(c)}{V_{n}(c)}V_{n}(\pi_x)\right)\,,
\end{align}  
with $c=-\frac{1+\delta^2}{1-\delta^2} $.
This result follows from the relation
\begin{align}
    \frac{V_{n+1}(c)}{V_{n}(c)}=\frac{\delta+1}{\delta-1}\,,
\end{align}
which can be derived using \eqref{eq:Uexp}. We conclude that the sequence $(Q_n)_n$ is obtained by doubling the sequence $(V_n)_n$.
\end{rmk}

\subsection{SSH model with non-vanishing chemical potential}

The construction explained in Remark \ref{rem:mar} is more general than the result used previously. Indeed, instead of using \eqref{eq:pi1}, one may define 
\begin{align}\label{eq:pi2}
  \pi_x=\frac{2}{1-\delta^2}(x-\mu^+)(x-\mu^-)-\frac{1+\delta^2}{1-\delta^2} \,,
\end{align}
and define the sequence $(Q_n)_n$ by
\begin{align}\label{eq:Qevenmu}
 &  Q_{2n}(x)= \frac{1+\delta}{2} U_{n}(\pi_x)+\frac{1-\delta}{2} U_{n-1}(\pi_x)\,,\\
 \label{eq:Qoddmu}
   & Q_{2n+1}(x)=(x-\mu^+)U_{n}(\pi_x)\,.  
\end{align}
Following the same computations as in the previous subsection, we find that the vector $\bold{Q}(x),$  whose components are $Q_0(x), Q_1(x), \dots, Q_{2N}(x)$ is an eigenvector of 
\begin{equation}\label{eq:alt}
 H=\left( \begin{array}{c c c c c c c}
    \mu^+& t^+ & \\
    t^+ & \mu^- & t^-\\
    0 &t^- & \mu^+ & t^+ \\
    &&& \ddots\\
    &&&t^+ & \mu^- & t^- \\
    &&&&t^- & \mu^+
    \end{array} \right)\,,
\end{equation}
with $x$ an eigenvalue chosen among 
\begin{align}
    \mu^+, \  x_k^\pm=\frac{\mu^++\mu^-\pm\sqrt{(\mu^+-\mu^-)^2+2(1+\delta^2)+2(1-\delta^2)\cos\left(\frac{k\pi}{N+1}\right)}}{2}\,,\quad (k=1,2,\dots, N)\,.
\end{align}
The parameters $\mu^\pm$ can be interpreted as a chemical potential in the SSH model, which can depend on the parity of the sites.

\section{Inhomogeneous SSH model from the doubling procedure}

\label{sec:SSS-doubling}
Inspired by the fact that the doubling procedure described in Remark \ref{rem:mar} applies to any sequence of orthogonal polynomials, we can use the previous construction to obtain inhomogeneous models \eqref{eq.HSSHinh} that generalize the SSH model and remain exactly solvable.
In this case, the tridiagonal matrix to diagonalize reads as 
\begin{equation}\label{eq:EigQinh}
 H^{(inh)}=\left( \begin{array}{c c c c c c c}
    0& \Jp_0 & \\
    \Jp_0 & 0 & \Jm_0 \\
    0 &\Jm_0 & 0 & \Jp_1 \\
    &&& \ddots\\
    &&&\Jp_{N-1} & 0 & \Jm_{N-1} \\
    &&&&\Jm_{N-1} & 0
    \end{array} \right)\,.
\end{equation}

We first describes explicitly the zero mode for this matrix. Next, we recall well-known properties of orthogonal polynomials necessary for the following construction. Using this framework, we propose an ansatz, based on the doubling procedure, for the eigenvectors of the inhomogeneous models. Finally, we establish the parameter constraints required to ensure the existence of these eigenvectors.

\subsection{Zero mode}

The matrix in \eqref{eq:EigQinh} can be interpreted as the single-particle Hamiltonian of an inhomogeneous SSH model which is represented in Fig.~\ref{fig:graph2}. The existence of a zero mode is a direct consequence of the chiral symmetry of the model (see e.g. \cite{asboth2016short,chiu2016classification}). One can check that the vector $\textbf{Q}^{(0)}$, whose components $Q_0^{(0)},Q_1^{(0)},\dots, Q_{2N}^{(0)}$ are defined by 
\begin{equation}\label{eq:zm}
    Q_{2n+1}^{(0)} =0\,,\quad (n=0,1,\dots, N-1), \qquad Q_{2n}^{(0)}=(-1)^n\prod_{i=0}^{n-1}\frac{t_{i}^+}{t_{i}^-}\,, \quad (n=0,1,\dots , N)\,,
\end{equation}
is an eigenvector of $H^{(inh)}$ with eigenvalue $0$. We will describe the other modes in Subsec. \ref{sec:SolInhSSH}. The existence of this zero mode is the main reason to focus the study on models with odd numbers of sites. We will also investigate a model with an even number of sites in Subsec. \ref{sec:SecondQRacahModel}. 

For the homogeneous SSH chain with $t^+_n=t^+$ and $t^-_n=t^-$, related to Chebyshev polynomials, the zero mode is localized at the end of the chain, as discussed in Sec.~\ref{sec:diaCh}. From Eq.~\eqref{eq:zm}, it is direct to conclude that the zero mode is localized on the last site of the chain for $t^+>t^-$, whereas it is localized on the first site for $t^+<t^-$. In the inhomogeneous case, we focus on functions $t^+_n$ which monotonically decrease along the chain, whereas functions $t^-_n$ monotonically increase. In this case, unlike for the homogeneous SSH chain, we expect the zero mode to be localized in regions of space where $t_n^+ \sim t_n^-$, thereby separating regions where $t_n^+>t_n^-$ and $t_n^+<t_n^-$, which correspond to different topological phases. This behavior is again a direct consequence of Eq.~\eqref{eq:zm}. Of course, the precise extension of the zero mode and the validity of this conjecture need to be verified in specific examples, which will be addressed in a forthcoming publication \cite{ssh_phys}.

\subsection{Orthogonal polynomials} 

Hypergeometric orthogonal polynomials, along with their $q$-analogs \cite{AskeyWilson, GasperRahman2004, Chiara, KoekoekLeskyetal2010}, form a central class of special functions. These families of polynomials are systematically organized in the ($q$-)Askey scheme and are characterized by a three-term recurrence relation in view of Favard's theorem. Moreover, they exhibit a bispectral property: they satisfy both a recurrence relation in the degree and a second-order differential or ($q$-)difference equation in the variable, which justifies their classification as classical polynomials. Understanding the detailed properties of these functions is valuable, as it enables their effective application in diverse mathematical and physical contexts. The families of orthogonal polynomials in the ($q$-)Askey scheme are defined in terms of ($q$-)hypergeometric functions \eqref{eq:q-hypergeom-func}.
For our purposes, we concentrate on the finite families within the ($q$-)Askey scheme, the most general of which are the ($q$-)Racah polynomials, and we consider only the properties necessary for the subsequent analysis.  

We denote by $P_n(x) = P_n(x; \rr)$, with $n = 0,1,\dots,N-1$ and $N$ a non-negative integer, the polynomials of a given finite family. Here, $\rr = (\rho_1, \rho_2, \dots)$ represents the set of parameters that define the family, including the integer $N$. These polynomials satisfy a 
three-term recurrence relation
\begin{align}
    xP_n(x)=A_n P_{n+1}(x)-(A_n+C_n)P_{n}(x)+C_nP_{n-1}(x)\,,\quad (n=0,1,\dots,N-1)\label{eq:recuP}
\end{align}
with the initial conditions $C_0=0,$ $A_{N-1}=0,$ $P_0(x)=1$.
The previous relations for $n=0,1,\dots, N-2$ allows us to define recursively the polynomials $P_1$, $P_2,\dots, P_{N-1}$. The last relation for $n=N-1$ reads 
$xP_{N-1}(x)=C_{N-1} (P_{N-2}(x)-P_{N-1}(x))$ and it provides a constraint for  $N$ possible values of $x$ found as the roots of this equation. We denote these values $(\lambda(x))_{x=0}^{N-1}$ and call it the grid associated to the polynomials $P_n$.
Assuming $A_{n}C_{n+1}>0$ (which we assume in the following), Favard's theorem (see e.g. \cite{Chiara}) guarantees that the sequence $(P_n)_n$ is orthogonal,
\begin{align}
    \sum_{x=0}^{N-1}\Omega(x)P_n(\lambda(x))P_m(\lambda(x))=\delta_{n,m} \cN_n\,,
\end{align}
with a certain weight function $\Omega(x)$ and norm $\cN_n$.
The finite families possess several important additional properties. One of these is known as \emph{Wilson duality}: for any polynomials $P_n(x;\rr)$, there exist polynomials $\mathcal{P}_i(x;\rr')$ such that
\begin{equation}
\mathcal{P}_n(x;\rr') = P_x(n;\rr).
\end{equation}
Additionally, they satisfy \emph{contiguity relations}, that is, equations of the form
\begin{align}
\lambda^{+}_{x; \rr}\, R_i(x; \rr)
&= \sum_{\epsilon \in \mathscr{S}} \Phi^{\epsilon,+}_{i}\, R_{i+\epsilon}(\ox; \orr)\,, 
\label{eq:cons1}
\end{align}
where $\ox$ denotes a shift of $x$, namely $\ox = x+\eta$ with $\eta \in \{0,+1,-1\}$, $\orr$ is a set of modified parameters, and $\mathscr{S}$ is a finite set of integers. In \cite{Contiguity25}, contiguity relations for the finite families of the Askey scheme have been classified, with $\mathscr{S}$ taken to be one of the sets $\{0,-1\}$, $\{0,-1,1\}$, or $\{0,-1,-2\}$. We note that some of these contiguity relations are commonly referred to as (forward or backward) shift relations \cite{KoekoekLeskyetal2010}.

\subsection{Solving the inhomogeneous SSH model}\label{sec:SolInhSSH}
In what follows, we change the normalization of the sequence $P_n$ in order to obtain a more convenient recurrence relation.
Defining the sequence $(R_n)_n$ as
\begin{align}\label{eq:RP}
    R_n=\epsilon^n \prod_{k=0}^{n-1}\frac{A_k}{\sqrt{A_kC_{k+1}}} P_{n}\,, \qquad (\epsilon=\pm 1)\,,
\end{align}
the three-term recurrence relation can be rewritten in the symmetric form
\begin{align}
\label{eq:3tr-Rn}
    xR_n(x)=\epsilon \sqrt{A_nC_{n+1}} R_{n+1}(x)-(A_n+C_n)R_{n}(x)+\epsilon \sqrt{A_{n-1}C_n}R_{n-1}(x)\,.
\end{align}
\paragraph{Ansatz for the form of the eigenvectors.}

Following the construction used for the Chebyshev polynomials (see Eqs.~\eqref{eq:Qeven} and \eqref{eq:Qodd}), we define a new sequence of polynomials $(Q_n)^{2N}_{n=0}$ from $(R_n)^{N-1}_{n=0}$ as follows: 
\begin{subequations}\label{eq:Qgen}
    \begin{align}\label{eq:Qeveng}
   &Q_{2n}(x)=\Jp_n R_{n}(\pi_x)+\Jm_{n-1} R_{n-1}(\pi_x)\,, &&(n=0,\dots,N)\,,\\
 \label{eq:Qoddg}
   & Q_{2n+1}(x)=xR_{n}(\pi_x)\,, &&(n=0,\dots,N-1)\,.
\end{align}
\end{subequations}
Here, we use the convention, $\Jp_{N}=0$, $\Jm_{-1}=0$, and 
\begin{align}
    \pi_x=\tau_2 x^2 +\tau_0\,,
\end{align}
where $\tau_{0},\tau_2$ are parameters that need to satisfy certain constraints (see below). The $Q_n(x)$ polynomials allow one to diagonalize the matrix $H^{(inh)}$ defined in Eq.~\eqref{eq:EigQinh} associated to an inhomogeneous SSH model, as we show in the following paragraph.

\paragraph{Inhomogeneous SSH models.}
If the coupling constants $t^\pm_n$ of the models  are given by the following relations in terms of the recurrence relation coefficients $A_n$, $C_n$ of a sequence of polynomials:
\begin{subequations}\label{eq:cons}
\begin{align}
\label{eq:cons0}
& t_n^\pm\neq 0\,,\quad (n=0,1,\dots,N-1),\\
\label{eq:cons1}
   & \epsilon\sqrt{A_nC_{n+1}}=\tau_2\Jm_n\Jp_{n+1}\,,\\
   \label{eq:cons2}
    &A_n+C_n+\tau_0=-\tau_2((\Jp_n)^2+(\Jm_n)^2)\,,
\end{align}    
\end{subequations}
then the vector $\bold{Q}(x)$, whose components $Q_0(x),Q_1(x),\dots, Q_{2N}(x)$ are defined by the previous ansatz \eqref{eq:Qgen}, diagonalizes the $(2N+1)\times (2N+1)$ matrix $H^{(inh)}$ \eqref{eq:EigQinh},
\begin{equation}\label{eq:EigQ}
H^{(inh)}\bold{Q}(x)=x\bold{Q}(x)\,.
\end{equation}
The associated eigenvalues $x$ are given by 
\begin{align}\label{eq:eigenvg}
x_k^\pm=\pm\sqrt{\frac{\lambda(k)-\tau_0}{\tau_2}}\,, \quad (k=0,1,\dots,N-1)\,,
\end{align}
where $\lambda(k)$ is the grid associated to the orthogonal polynomials $(P_n(x))_{n=0}^{N-1}$. We recall that $H^{(inh)}$ has  also a vanishing eigenvalue with the eigenvector given by \eqref{eq:zm}.

The remaining part of this paragraph is devoted to the proof of these results.
The components of the eigenvalue problem \eqref{eq:EigQ} read
\begin{align}
     &t_{n-1}^-Q_{2n-1}(x)+t_{n}^+Q_{2n+1}(x)=xQ_{2n}(x)\,,&& (n=0,1,\dots,N)\,,\label{eq:Jmm}\\
    &\Jp_n Q_{2n}(x)+\Jm_n Q_{2n+2}(x)=x Q_{2n+1}(x)\,,&& (n=0,1,\dots,N-1)\,.\label{eq:Jpp}
\end{align}
Relation \eqref{eq:Jmm} is a direct consequence of the definitions \eqref{eq:Qeveng} and \eqref{eq:Qoddg}. 

To prove Eq.~\eqref{eq:Jpp},
let us compute 
\begin{align}
   xQ_{2n+1}(x)&=x^2R_{n}(\pi_x)=\frac{1}{\tau_2}(\tau_2 x^2 +\tau_0)R_{n}(\pi_x)-\frac{\tau_0}{\tau_2} R_{n}(\pi_x)\nonumber\\
   &=\frac{\epsilon}{\tau_2} \sqrt{A_nC_{n+1}}\; R_{n+1}(\pi_x)-\frac{1}{\tau_2}(A_n+C_n+\tau_0)R_{n}(\pi_x)+\frac{\epsilon}{\tau_2}  \sqrt{A_{n-1}C_n}\;R_{n-1}(\pi_x)\,.\label{eq:ret}
\end{align}
The last equality is obtained using the three-term recurrence relation for $R_n$. Using the constraints \eqref{eq:cons1} and \eqref{eq:cons2},
one obtains 
\begin{align}
    xQ_{2n+1}(x)&=\Jm_n\Jp_{n+1}R_{n+1}(\pi_x)+((\Jp_n)^2+(\Jm_n)^2)R_{n}(\pi_x)+\Jm_{n-1}\Jp_{n}R_{n-1}(x)\nonumber\\
   &= \Jm_n Q_{2n+2}(x)+\Jp_n Q_{2n}(x)\,,
\end{align}
which proves  \eqref{eq:Jpp}.
As mentioned after \eqref{eq:recuP},  the three-term recurrence relation used for \eqref{eq:ret} is valid only if $\pi_x$ is on the grid of the associated polynomials \textit{i.e.} $\pi_x=\lambda(k)$ ($k=0,1,\dots, N-1$). The eigenvalues \eqref{eq:eigenvg} are then obtained when solving $\pi_x=\lambda_k$. 
\paragraph{Finding the eigenvalues.}
We already showed that $x=0$ is an eigenvalue with eigenvector $\textbf{Q}^{(0)}$ defined by \eqref{eq:zm}. The vectors $\textbf{Q}(x)$ are eigenvectors if $x=x_k$ for some $k=0,\cdots, N-1$. Indeed, equation \eqref{eq:Jmm} is satisfied for all $x$ and all $n=0\cdots N$. The only issue appears in equation  \eqref{eq:Jpp} for $n=N-1$. In fact, we had to use the recurrence relation for the family $R_n$ which is valid only if $\pi_x$ is on the grid $(\lambda(k))$. Solving $\pi_x=\lambda(k)$ gives the  eigenvalues $x_k^\pm$ \eqref{eq:eigenvg}. 

\paragraph{Trivial solution.}
At first glance, the following solution 
\begin{equation}
        \epsilon=\tau_2=1\,,\quad \tau_0=0\,,\quad t_n^+=\sqrt{C_n},\quad t^-_n=\sqrt{A_n},
\end{equation}
seems to be valid for any finite family of orthogonal polynomials. However, the constraint $t^\pm_n\neq 0$ is not satisfied since $t_{N-1}^-=A_{N-1}=0=C_0=t_0^+$. Nevertheless, we can restrict ourselves to the subsystem made only of the sites $1,\ldots,2N-1$ (we do not consider the uncoupled sites $0$ and $2N$) and adapt the proof of the previous proposition to get the solution of this model. Fortunately, we do not need to study these restricted models since we can show that these can be obtained through the doubling of orthogonal polynomials. 

\paragraph{Homogeneous SSH model.}
To recover the SSH model described in Sec.~\ref{sec:diaCh}, we set 
\begin{align}
    \epsilon=+1\,,\quad \tau_0=-2-2\frac{1+\delta^2}{1-\delta^2}\,,\quad \tau_2=\frac{4}{1-\delta^2}\,,\quad \Jpm_n=\frac{1\pm \delta}{2}\,,\quad A_n=C_n=1\,,
\end{align}
and
\begin{align}\label{eq:RU}
    R_n(x)=U_n\left(\frac{x+2}{2}\right)\,.
\end{align}
Let us point out that, due to the affine transformation in the parameter of \eqref{eq:RU}, it is necessary to make the same affine transformation for the function $\pi_x$ to compare with the one given by \eqref{eq:pi1}: \begin{align}
\frac{\tau_0+\tau_2 x^2+ 2}{2}=\frac{2}{1-\delta^2}x^2-\frac{1+\delta^2}{1-\delta^2}\,.
\end{align}

\subsection{Connection with the doubling procedure} The observation regarding the usual SSH model made in remark \ref{rem:mar} remains valid for 
the general construction of the sequence $(Q_n)_n$ given in this section: $(Q_n)_n$ can be seen as the doubling procedure introduced in \cite{MARCELLAN1997}.
Indeed, let us introduce the sequence of polynomials $(V_n)_n$ defined by 
\begin{align}
    V_n(x)=t_n^+R_n(x)+t_{n-1}^-R_{n-1}(x)\,.
\end{align}
It is straightforward to observe that $Q_{2n}(x)=V_n(\pi_x)$. Moreover, we verify that
  \begin{equation}
  \label{eq:Qoddf1}
   Q_{2n+1}(x)=\frac{\tau_2t_n^- x}{\pi_x-\tau_0}\left(V_{n+1}(\pi_x)+\frac{t_n^+}{t_n^-} V_{n}(\pi_x)\right)\,.
\end{equation}  
This follows from the definition of $V_n(x)$ and the constrains \eqref{eq:cons1}, \eqref{eq:cons2}. Indeed, the right hand side of \eqref{eq:Qoddf1} is equal to
\begin{align}
 &\frac{\tau_2t_n^- x}{\pi_x-\tau_0}\left(t_{n+1}^+R_{n+1}(\pi_x)+\left(t_n^- +\frac{(t_n^+)^2}{t_n^-}\right)R_n(\pi_x)+\frac{t_n^+}{t_n^-} t_{n-1}^-R_{n-1}(\pi_x) \right)\nonumber\\
 &=\frac{\tau_2t_n^-x}{\pi_x-\tau_0}\left(\frac{\epsilon \sqrt{A_nC_{n+1}}}{\tau_2t_n^-}R_{n+1}(\pi_x)-\frac{(A_n+C_n+\tau_0)}{\tau_2t_n^-}R_n(\pi_x)+\frac{\epsilon \sqrt{A_{n-1}C_{n}}}{\tau_2t_n^-}R_{n+1}(\pi_x) \right)\nonumber\\
 &=\frac{\tau_2t_n^- x}{\pi_x-\tau_0}\ \frac{\pi_x-\tau_0}{\tau_2 t_n^-}R_n(\pi_x)=Q_{2n+1}(x),
\end{align}
which is the left hand side of \eqref{eq:Qoddf1}.
In addition to that, we have the following property:
\begin{align}
  V_n(\tau_0) t_n^++V_{n+1}(\tau_0)t_n^-&= (t_n^+)^2R_n(\tau_0)+t_n^+t_{n-1}^-R_{n-1}(\tau_0)+ t_n^-t_{n+1}^+R_{n+1}(\tau_0)+(t_{n}^-)^2R_{n}(\tau_0)\nonumber\\
  &=\frac{1}{\tau_2}\left(\epsilon\sqrt{A_nC_{n+1}} R_{n+1}(\tau_0)-(A_n+C_n+\tau_0)R_n(\tau_0)+\epsilon\sqrt{A_{n-1}C_{n}}R_{n-1}(\tau_0)\right)\nonumber\\
  &=\frac{1}{\tau_2}(\tau_0R_n(\tau_0)-\tau_0R_n(\tau_0))
  =0,
\end{align}
which is equivalent to  $\frac{t_n^+}{t_n^-}=-\frac{V_{n+1}(\tau_0)}{V_n(\tau_0)}$. Finally, relation \eqref{eq:Qoddf1} becomes
 \begin{equation}
  \label{eq:christoffel-general}
   Q_{2n+1}(x)=\frac{\tau_2t_n^- x}{\pi_x-\tau_0}\left(V_{n+1}(\pi_x)-\frac{V_{n+1}(\tau_0)}{V_n(\tau_0)}  V_{n}(\pi_x)\right)\,,
   \end{equation}
   which is the Christoffel transformation used in \cite{MARCELLAN1997}\footnote{There are small modifications with respect to \cite{MARCELLAN1997} due to the fact that the polynomials considered here are not monic and that the coefficient of $x^2$ in $\pi_x$ is not 1.}.

In the following, we use the above construction with well-known finite sequences of orthogonal polynomials from the ($q$-)Askey scheme (see e.g. \cite{KoekoekLeskyetal2010}) to provide examples of inhomogeneous generalizations of the SSH model, and hence construct interesting solvable physical models.
The results for each family of polynomials discussed in this work, including their corresponding couplings and eigenvalues, are summarized in the following table:
\begin{table}[h]
\centering
\small
\renewcommand{\arraystretch}{2}
\begin{tabular}{|l|c|c|c|c|c|}
  \hline
  Polynomials & $t^+_n$ & $t^-_n$ & $\tau_2$ & $\tau_0$ & Eigenvalues \\
  \hline\hline
  Chebyshev
  & $(1+\delta)/2$
  & $(1-\delta)/2$
  & $\frac{4}{1-\delta^2}$
  & $-2-2\frac{1+\delta^2}{1-\delta^2}$
  & $0,\ \pm\sqrt{\frac{1-\delta^2}{2}\cos\!\left(\frac{k\pi}{N+1}\right)
    +\frac{1+\delta^2}{2}}$ \\
  \hline
  Krawtchouk
  & $\sqrt{p(N-n)}$
  & $\sqrt{(1-p)(n+1)}$
  & $-1$
  & $1$
  & $0,\ \pm\sqrt{k+1}$ \\
  \hline
  $q$-Racah
  & \eqref{eq:tnpqRacah}
  & \eqref{eq:tnmqRacah}
  & $-1$
  & $-(1-\delta)(1-q^{-N})$
  & $0,\ \pm\sqrt{(1-q^{k-N})(\delta-q^{-k})}$ \\
  \hline
\end{tabular}

\caption{Summary of the different SSH models studied in this paper.}
\label{Tbl:summary}
\end{table}

\section{Inhomogeneous SSH model associated to Krawtchouk polynomials}
\label{sec:Krawtchouk}

\subsection{Properties of Krawtchouk polynomials}
In this section, we give a solution to the constraints \eqref{eq:cons} associated to the Krawtchouk polynomials. 
These polynomials are defined as follows in terms of the ${}_2F_1 $ hypergeometric function:
\begin{align}
    K_n(x;p,N-1)={}_2F_1 \left({{ -n,\;-x}\atop
{-N+1}}\;\Bigg\vert  \frac{1}{p}\right)\,,\quad (0<p<1\,,\quad n=0,1,\dots,N-1)\,,
\end{align}
and satisfy the three-term recurrence relation
\begin{align}
  &  -x K_n(x;p,N-1)=\nonumber\\
   & \qquad A_n^{(K)}K_{n+1}(x;p,N-1)-(A^{(K)}_n+C^{(K)}_n)K_{n}(x;p,N-1)+C^{(K)}_n K_{n-1}(x;p,N-1)\,,
\end{align}
with
\begin{align}
    A_n^{(K)}=p(N-n-1)\,,\quad  C_n^{(K)}=n(1-p)\,.
\end{align}
Equivalently, they also satisfy the orthogonality relation 
\begin{multline}\label{eq:OrthoKrawtchouk}
    \sum_{x=0}^{N-1} \binom{N-1}{x}p^x(1-p)^{n-1-x}K_n(x;p,N-1)K_m(x;p,N-1)=\frac{(-1)^nn!}{(-N+1)_n}\left(\frac{1-p}{p}\right)^n\delta_{n,m}.
\end{multline}
As discussed in the previous section, in order to obtain a more convenient recurrence relation, we define renormalized Krawtchouk polynomials, for $\epsilon=\pm 1$, as
\begin{align}
    R^{(K)}_n(x)=\epsilon^n \sqrt{\frac{p^n}{(1-p)^n}\binom{N-1}{n}} K_n(-x;p,N-1)\,,
\end{align}
which satisfy
\begin{align}
    xR^{(K)}_n(x)=&\epsilon \sqrt{A_{n}^{(K)}C_{n+1}^{(K)}}\;R^{(K)}_{n+1}(x)-(A_n^{(K)}+C_n^{(K)})R^{(K)}_{n}(x)+\epsilon \sqrt{A_{n-1}^{(K)}C_{n}^{(K)}}\;R^{(K)}_{n-1}(x)\,.
\end{align}
The grid associated to this sequence of polynomials is $\lambda(k)=-k$, ($k=0,1,\dots, N-1$).

\subsection{SSH model of Krawtchouk type}

For this choice of polynomials, the constraints \eqref{eq:cons1} and \eqref{eq:cons2} read
\begin{subequations}
\begin{align}\label{eq:consK1}
   & \epsilon\sqrt{p(1-p)(N-n-1)(n+1)}=\tau_2\Jm_n\Jp_{n+1}\,,\\
   \label{eq:consK2}
    &-(p(N-n-1)+n(1-p)+\tau_0)=\tau_2((\Jp_n)^2+(\Jm_n)^2)\,.
\end{align}
\end{subequations}
A solution to these constraints is 
\begin{align}
    \epsilon=\tau_2=-1\,,\quad \tau_0=1\,,\quad \Jp_n=\sqrt{p(N-n)}\,,\quad \Jm_n=\sqrt{(1-p)(n
    +1)}\,.
\end{align}
With these choices, results of Sec.~\ref{sec:SSS-doubling}
allow us to diagonalize exactly the following  $(2N+1) \times (2N+1)$ matrix, 
\begin{equation}
 H=\left( \begin{array}{c c c c c c c}
    0& u_0\sqrt{1+\delta} & \\
    u_0 \sqrt{1+\delta} & 0 & u_{N-1}\sqrt{1-\delta} \\
     &u_{N-1}\sqrt{1-\delta} & 0 & u_1\sqrt{1+\delta} \\
    &&& \ddots\\
    &&&u_{N-1}\sqrt{1+\delta} & 0 & u_0\sqrt{1-\delta} \\
    &&&&u_0\sqrt{1-\delta} & 0
    \end{array} \right)\,.
\end{equation}
Here, $\delta=2p-1$ and $u_i=\sqrt{\frac{N-i}{2}}$.
The spectrum is given by
\begin{equation}
    0, x_k^\pm=\pm\sqrt{k+1},\qquad (k=0,1,\dots, N-1).
\end{equation}
Using results of Sec.~\ref{sec:SSS-doubling} and after manipulating the Krawtchouk polynomials, the components of the eigenvectors $\bold{Q}(x_i^\pm)$ ($i=0,1,\dots,N-1$) are 
\begin{subequations}
    \begin{align}
&Q_{2n}(x_k^\pm)=(-1)^n\sqrt{\frac{Np^{n+1}}{(1-p)^n}\binom{N}{n}} \;K_n(k+1;p,N)\,,\label{eq:Qes}\\
 &   Q_{2n+1}(x_k^\pm)=\pm(-1)^n \sqrt{\frac{(k+1)p^n}{(1-p)^n}\binom{N-1}{n}}\; K_n(k;p,N-1)\,.
\end{align}
\end{subequations}
To prove relation \eqref{eq:Qes}, we used the contiguity relation $(KI)$ of Ref.~\cite{Contiguity25} (also known as the backward shift relation of the Krawtchouk polynomials \cite{KoekoekLeskyetal2010} up to self duality), \textit{i.e.},
\begin{align}\label{eq:dbs}
    (1-p)nK_{n-1}(k;p,N)-p(N+1-n)K_n(k;p,N) =-p(N+1)K_n(k+1;p,N+1)\,.
\end{align}
The components of the eigenvector $\bold{Q}^{(0)}$, associated to the vanishing eigenvalue (or zero mode), are:
\begin{align}
    Q_{2n}^{(0)}=(-1)^n\sqrt{\frac{p^{n}}{(1-p)^n}\binom{N}{n}} \,, \quad Q_{2n+1}^{(0)}=0\,.
\end{align}

The zero mode amplitude $ \left(Q_{2n}^{(0)}\right)^2$ is maximal close to $n \sim pN$, which corresponds to the position in the chain where $t_n^+ \sim t_n^- \sim \sqrt{Np(1-p)}$ for the Krawtchouk case. Because $t_n^+$ (resp. $t_n^-$) is monotonically decreasing (increasing) with $n$, the zero mode indeed separates regions where $t_n^+>t_n^-$ and $t_n^+<t_n^-$, in agreement with our general discussion of Sec.~\ref{sec:SSS-doubling}.

The eigenvectors $\bold{Q}(x_k^\pm)$, $\bold{Q}^{(0)}$ are orthogonal, with norms
\begin{align}
\label{eq:norms-krawt}
    \bold{Q}(x_k^\pm)^t \ \bold{Q}(x_k^\pm)=\frac{2(k+1)(1-p)^{i-N+1}}{p^i\binom{N-1}{i}}\,,\qquad
     {\bold{Q}^{(0)}}^t \ \bold{Q}^{(0)}=\frac{1}{(1-p)^{N}}\,,
\end{align}
for $k=0,1,\dots,N-1$. To prove \eqref{eq:norms-krawt}, we use the self-duality property of the Krawtchouk polynomials, namely \( K_n(x; p, N) = K_x(n; p, N) \), which is obvious from their hypergeometric representation, together with the orthogonality relation \eqref{eq:OrthoKrawtchouk}.

\section{Inhomogeneous SSH model associated to $q$-Racah polynomials}
\label{SSH-qRacah}
In the case of the Krawtchouk polynomials discussed in the previous section, a significant simplification arises in the computation \eqref{eq:Qes} of $Q_{2n}(x)$, since it can be expressed as a single Krawtchouk polynomial. The relations of type \eqref{eq:dbs} used to achieve this simplification are called contiguity relations, and analogous relations exist for other finite sequences in the ($q$-)Askey scheme. This construction is now extended to the $q$-Racah polynomials, making use of the classification of contiguity relations provided in \cite{Contiguity25}. In particular, in the following we define two different solvable models associated to these polynomials.

\subsection{Properties of the $q$-Racah polynomials}

The $q$-Racah polynomials $P^{(qR)}_n$ are defined in terms of $q$-hypergeometric functions as follows, for $N$ a nonnegative integer
\begin{align}\label{eq:qRacah}
P^{(qR)}_n(\lambda_{x};\rr)={}_4\phi_3 \left({{q^{-n},\; \alpha\beta q^{n+1}, \;q^{-x},\;\gamma\delta  q^{x+1}}\atop
{\alpha q,\; \beta \delta q,\;\gamma q }}\;\Bigg\vert \; q;q\right)\,,\quad (n=0,1,\dots,N)\,,
\end{align}
where $\lambda_{x}=q^{-x}+\gamma\delta q^{x+1},$ $\rr=\alpha,\beta,\gamma,\delta$ and one of the following truncation conditions is imposed:
\begin{equation}
    \alpha q=q^{-N},\quad \beta \delta q= q^{-N},\quad\textrm{or} \quad \gamma q=q^{-N}.
\end{equation} 
The $q$-Racah polynomials enjoy several remarkable properties, including orthogonality, bispectrality and duality \cite{KoekoekLeskyetal2010}. In particular, they satisfy the orthogonality relation 
\begin{align}
\label{eq:orth-qracah}
    \sum_{x=0}^N P_n^{(qR)}(\lambda_x;\rr)P_m^{(qR)}(\lambda_x;\rr)w(x;\rr)=\delta_{n,m}h_{n,\rr}\,,
\end{align}
where the weight function and the norm are given by
\begin{align}
w(x;\rr)&=\frac{(\alpha q,\beta\delta q,\gamma q,\gamma \delta q;q)_x (1-\gamma\delta q^{2x+1})}{(q,\delta\gamma\alpha^{-1} q,\beta^{-1} \gamma q,\delta q;q)_x(\alpha\beta q)^x(1-\gamma\delta q)}\,,\\
h_{n,\rr}&=\frac{(\alpha^{-1}\beta^{-1} \gamma, \alpha^{-1}\delta, \beta^{-1},\gamma\delta q^2;q)_\infty}{(\alpha^{-1}\beta^{-1}q^{-1},\alpha^{-1}\gamma\delta q, \beta^{-1}\gamma q,\delta q;q)_\infty}\frac{(1-\alpha\beta q)(\gamma \delta q)^n}{(1-\alpha\beta q^{2n+1})}\frac{(q,\alpha\beta\gamma^{-1} q,\alpha \delta^{-1} q,\beta q;q)_n}{(\alpha q, \alpha\beta q, \beta\delta q, \gamma q;q)_n}\,.
\end{align}
The polynomials $P_n^{(qR)}(\lambda_x;\rr)$ satisfy the following three-term recurrence relation:
\begin{align}
-(1-q^{-x})(1-\gamma\delta &q^{x+1})P_n^{(qR)}(\lambda_x;\rr)=\nonumber\\
&\qquad A^{(qR)}_{n,\rr}\; P^{(qR)}_{n+1}(\lambda_x; \rr)
-(A^{(qR)}_{n,\rr}+C^{(qR)}_{n,\rr})\;P^{(qR)}_n(\lambda_x; \rr)+C^{(qR)}_{n,\rr}\; P^{(qR)}_{n-1}(\lambda_x; \rr)\,,
\end{align}
where 
\begin{subequations}
\begin{align}
\label{eq:coeff-3t-qracah}
 &   A^{(qR)}_{n,\rr}=\frac{(1-\gamma q^{n+1})(1-\alpha q^{n+1})(1-\alpha\beta q^{n+1})(1-\beta\delta q^{n+1})}{(1-\alpha\beta q^{2n+1})(1-\alpha\beta q^{2n+2})}\,,\\
  &  C^{(qR)}_{n,\rr}=\frac{q(1-q^{n})(1-\beta q^{n})(\gamma-\alpha\beta q^{n})(\delta-\alpha q^n)}{(1-\alpha\beta q^{2n})(1-\alpha\beta q^{2n+1})}\,.
\end{align}
\end{subequations} 
Moreover, the polynomials $P_n^{(qR)}(\lambda_x;\rr)$ satisfy the duality relation
\begin{equation}
\label{eq:duality-qracah}
P_n^{(qR)}(\lambda_x;\rr)=P_x^{(qR)}(\lambda_n^d;\rr^d)\,,\quad (n,x=0,1,\dots N)\,,
\end{equation}
where $\lambda_n^d= q^{-n}+\alpha\beta q^{n+1}$ and $\rr^d=\gamma,\delta,\alpha,\beta.$
Furthermore, the following identity holds  
\begin{equation}
\label{eq:dual-weight-prod}
w(n;\rr^d)=\prod_{k=0}^{n-1}\frac{A^{(qR)}_{k,\rr}}{C^{(qR)}_{k+1,\rr}}\,.
\end{equation}
Consequently, the normalization appearing in \eqref{eq:RP}
is related to the weight as
\begin{align}
    \epsilon^n \prod_{k=0}^{n-1}\frac{A^{(qR)}_{k,\rr}}{\sqrt{A^{(qR)}_{k,\rr}C^{(qR)}_{k+1,\rr}}}=\epsilon^n(-1)^{\nu_n} \sqrt{w(n;\rr^d)},
\end{align}
where $\nu_n$ denotes the number of indices $k\in\{0,1,\ldots,N\}$ such that $A^{(qR)}_{k,\rr}$ is negative.

\subsection{First exactly solvable model}
\label{sec:RacahI}

In this subsection, we consider the $q$-Racah polynomials with parameters $\rr = \alpha, q\beta,q^{-N },\delta/q$, and renormalize them as follows:
\begin{equation}
R^{(qR)}_n(x,\rr)=\epsilon^n(-1)^{\nu_n}\, \sqrt{w(n;\rr^d)} P^{(qR)}_{n}(\lambda_x,\rr)\,,
\end{equation} 
where $ \epsilon=\pm1$ and $\rr^d=q^{-N},\delta/q,\alpha,q\beta.$ The orthogonality grid for this sequence of polynomials is
\begin{equation}
    \lambda_{k,\rr}=-(1-q^{-k})(1-\delta q^{k-N}),\qquad (k=0,1,\dots, N-1).
\end{equation}
\begin{prop}\label{pr:qracah}
A solution to the constraints \eqref{eq:cons} is given by 
\begin{subequations}
\begin{align}
\tau_0&=-\left(1-\delta  \right)(1-q^{-N})\,,\quad\tau_2=-1\,,\quad \epsilon=-1\,,\\
t_n^+&=\sqrt{\frac{(1-q^{n-N})(1-\alpha\beta q^{n+1})(1-\beta q^{n+1})(\delta-\alpha q^{n+1})}{(1-\alpha \beta q^{2n+1})(1-\alpha\beta q^{2n+2})}}\,,\label{eq:tnpqRacah}\\
t_n^-&=\sqrt{\frac{(1-\beta\delta q^{n+1})(1-q^{n+1})(1-\alpha\beta q^{n+N+2})(1-\alpha q^{n+1})}{q^{N} (1-\alpha\beta q^{2n+2})(1-\alpha\beta q^{2n+3})}}\,.\label{eq:tnmqRacah}
\end{align}
\end{subequations}
\end{prop}
\begin{proof}
The proof follows directly from a straightforward calculation using the expressions for \(\tau_0\), \(\tau_2\), \(t_n^+\), and \(t_n^-\).
\end{proof}
With these solutions, the general results obtained in Sec.~\ref{sec:SSS-doubling} allow one to exactly diagonalize the corresponding $(2N + 1) \times (2N + 1)$ matrix given in \eqref{eq:EigQ}, whose spectrum is 
\begin{equation}
    0, \ x_k^\pm=\pm\sqrt{(1-q^{k-N})(\delta -q^{-k})}, \qquad (k=0,1,\dots, N-1).
\end{equation}
Using results of Sec.~\ref{sec:SSS-doubling} and performing suitable manipulations of the $q$-Racah polynomials, the components of the eigenvectors $\boldsymbol{Q}(x_k^\pm)$, for $k = 0, 1, \dots, N-1$, are given by
\begin{subequations}
    \begin{align}
&Q_{2n}(x_k^\pm)=\epsilon^n(-1)^{\nu_n}\sqrt{\frac{(1-\beta q)(\delta-\alpha q)(1-q^{-N})w(n,\orr^d)}{1-\alpha\beta q^2}}P_n^{(qR)}(\lambda_k;\orr)\,,&& (n=0,\ldots,N)\,,\label{eq:Q-racah-1st-case}\\
 &   Q_{2n+1}(x_k^\pm)=\pm\epsilon^n(-1)^{\nu_n}\sqrt{(1-q^{k-N})(\delta -q^{-k})w(n,\rr^d)}P_n^{(qR)}(\lambda_k;\rr)\,,&&(n=0,\ldots,N-1)\,.
\end{align}
\end{subequations}
The parameters are specified as follows:
\begin{equation}
    \rr = \alpha, q\beta, q^{- N}, \delta/q,\quad \rr^d=q^{-N},\delta/q,\alpha,q\beta,\quad\orr=\alpha,\beta,q^{-N-1},\delta\quad \text{and}\quad \orr^d=q^{-N-1},\delta,\alpha,\beta. 
\end{equation} To establish the relation \eqref{eq:Q-racah-1st-case}, we use the following contiguity relation of the $q$-Racah polynomials (relation $(qRI)$ of Ref.~\cite{Contiguity25}):
\begin{align}\label{eq:qRI}
P^{(qR)}_n(\lambda_x;\orr)&=\frac{(1- q^{n-N})(1-\alpha\beta q^{n+1})}{(1-q^{-N})(1-\alpha\beta q^{2n+1})}P^{(qR)}_n(\lambda_x,\rr)+\frac{(1-q^n)(1-\alpha \beta q^{N+n+1})}{(1-q^{N})(1-\alpha\beta q^{2n+1})}P^{(qR)}_{n-1}(\lambda_x,\rr).
\end{align}
The components of the eigenvector $\bold{Q}^{(0)}$, corresponding to the vanishing eigenvalue, are given by
\begin{align}
    Q_{2n}^{(0)}=(-1)^n\sqrt{\frac{(q^N\delta)^n(1-\alpha\beta q^{2n+1})(q^{-N},\alpha\beta q,\beta q,\alpha q/\delta;q)_n}{(1-\alpha\beta q)(q,\beta\delta q,\alpha \beta q^{N+2},\alpha q;q)_n }} \,, \quad Q_{2n+1}^{(0)}=0\,.
\end{align}
The eigenvectors $\bold{Q}(x_k^\pm)$, $\bold{Q}^{(0)}$ are orthogonal, with norms, for $k=0,1,\dots,N-1$:
\begin{align}
    \bold{Q}(x_k^\pm)^t \ \bold{Q}(x_k^\pm)
  &=2(1-q^{k-N})(\delta-q^{-k})h_{k,\rr^d},\label{eq:norms-racah-c1}\\
 \qquad   {\bold{Q}^{(0)}}^t \ \bold{Q}^{(0)}&=\frac{(\delta,\alpha\beta q^2;q)_N}{(\alpha q,\beta \delta q;q)_N}\label{eq:norms0-racah-c1}\,.
\end{align}
The relations \eqref{eq:norms-racah-c1} and \eqref{eq:norms0-racah-c1} follow from the duality property \eqref{eq:duality-qracah} of the $q$-Racah polynomials.

Finally, we remark that the contiguity relation $(qRIII)$ given in \cite{Contiguity25} yields the same result as above, up to the following reparameterization:
\begin{align}
    \beta \to \alpha, \qquad \delta\to\frac{1}{\delta}.
\end{align}

\subsection{Second exactly solvable model}\label{sec:SecondQRacahModel}

Additional contiguity relations for $q$-Racah polynomials were derived in \cite{Contiguity25}. Some of these relations do not alter the value of $N$, in contrast to those used previously (see \eqref{eq:dbs} or \eqref{eq:qRI}). By employing such relations, alternative solutions to the constraints can be obtained but they apply to a chain with an even number of sites and a slight modification of results of Sec.~\ref{sec:SSS-doubling} is needed. 

Consider now $q$-Racah polynomials with parameters $\rr=\alpha,q\beta,q^{-N},\delta$, and renormalize them as follows:
\begin{equation}
R^{(qR)}_n(x,\rr)=\epsilon^n\, \sqrt{w(n;\rr^d)} P^{(qR)}_{n}(\lambda_x,\rr)\,,
\end{equation} 
where $ \epsilon=\pm1$ and $\rr^d=q^{-N},\delta,\alpha,q\beta.$
The grid for this sequence of polynomials is \begin{equation}
    \lambda_{k,\rr}=-(1-q^{-k})(1-\delta q^{k-N+1}),\qquad (k=0,1,\dots, N-1).
    \end{equation}
\begin{prop}
A solution to the constraints \eqref{eq:cons1} and \eqref{eq:cons2} is given by 
\begin{subequations}
\begin{align*}
\tau_0&=-\left(1-\beta^{-1}  q^{-N}\right) \left(1-q\beta  \delta  \right)\,,\quad\tau_2=-\beta^{-1}q^{-N}\,,\quad\epsilon=1\,,\\
t_n^+&=\sqrt{\frac{(1-\beta q^{n+1})(1-\alpha\beta q^{n+1})(1-\alpha\beta q^{n+N+1})(1-\beta\delta q^{n+1})}{(1-\alpha\beta q^{2n+1})(1-\alpha\beta q^{2n+2})}}\,,\\
t_n^-&=- q\beta \sqrt{\frac{q^{N-1}(1-q^{n-N+1})(1-\alpha q^{n+1})(1-q^{n+1})(\delta-\alpha q^{n+1})}{(1-\alpha\beta q^{2n+2})(1-\alpha\beta q^{2n+3})}}.
\end{align*}
\end{subequations}
\end{prop}
\begin{proof}
The proof follows directly from a straightforward calculation using the expressions for \(\tau_0\), \(\tau_2\), \(t_n^+\), and \(t_n^-\).
\end{proof}
In this case, $t_{N-1}^-=0$. We can eliminate the last row and column of the \((2N+1) \times (2N+1)\) matrix and consider the resulting subsystem. By adapting the proof given in Sec.~\ref{sec:SSS-doubling}, we can show that the spectrum of the corresponding $2N   \times 2N $ matrix is given by
\begin{equation}
     x_k^\pm=\pm\sqrt{(1-\delta\beta q^{k+1})(1-\beta q^{N-k})}, \qquad (k=0,1,\dots, N-1), 
\end{equation}
and the components of the eigenvectors $\boldsymbol{Q}(x_k^\pm)$, for $k = 0, 1, \dots, N-1$, are given by
\begin{subequations}
    \begin{align}
&Q_{2n}(x_k^\pm)=(-1)^n\sqrt{\frac{(1-\beta q)(1-\alpha\beta q^{N+1})(1-\beta\delta q)w(n;\orr^d)}{1-\alpha\beta q^2}}P_n^{(qR)}(\lambda_k;\orr),&&(n=0,\ldots, N-1) \label{eq:Q-racah-2nd-case}\\
 &   Q_{2n+1}(x_k^\pm)=\pm(-1)^n\sqrt{(1-\delta\beta q^{k+1})(1-\beta q^{N-k})w(n;\rr^d)}P_n^{(qR)}(\lambda_k;\rr),&&(n=0,\ldots, N-1)\,.
\end{align}
\end{subequations} 
The parameters are specified as follows:
\begin{equation}
\rr=\alpha,q\beta,q^{-N},\delta,\quad \orr=\alpha,\beta,q^{-N},\delta,\quad\rr^d=q^{-N},\delta,\alpha,q\beta\quad \text{and}\quad \orr^d=q^{-N},\delta, \alpha,\beta.   
\end{equation}
 To establish the relation \eqref{eq:Q-racah-2nd-case}, we employ the following contiguity relation of the $q$-Racah polynomials (see $(qRII)$ in \cite{Contiguity25}):
\begin{align}
P^{(qR)}_n(\lambda_x;\orr)&=\frac{(1-\alpha\beta q^{n+1})(1-\beta\delta q^{n+1})}{(1-\beta\delta)(1-\alpha\beta q^{2n+1})}P^{(qR)}_n(\lambda_x,\rr)-\frac{\beta (1-q^n)(\delta-\alpha q^n)}{(1-\beta\delta )(1-\alpha\beta q^{2n+1})}P^{(qR)}_{n-1}(\lambda_x,\rr).
\end{align}
We note that there is no shift in the parameter $N-1$ when comparing $\rr$ and $\orr$. The eigenvectors $\bold{Q}(x_i^\pm)$ are orthogonal, with norms:
\begin{align}
\label{eq:norms-qracah-c2}
    \bold{Q}(x_k^\pm)^t \ \bold{Q}(x_k^\pm)=2(1-\delta\beta q^{k+1})(1-\beta q^{N-k})h_{k,\rr^d}\,,\quad (k=0,1,\dots,N-1)\,.
\end{align}
The relations \eqref{eq:norms-qracah-c2} follow from the duality property \eqref{eq:duality-qracah} of the $q$-Racah polynomials.

Finally, we remark that the contiguity  relation $(qRIV)$ given in \cite{Contiguity25} yields the same result as above, up to the following reparametrization:
\begin{align}
    \beta \to \alpha, \qquad \gamma\to\frac{1}{\gamma}.
\end{align}


\section{Conclusion} 

Summing up, we have shown that the eigenvalues and eigenvectors of the standard SSH model can be elegantly derived through a connection with Chebyshev polynomials and the doubling method for polynomial sequences. Exploiting this analytical framework, we introduced new  exactly solvable models that extend the conventional SSH model by allowing for inhomogeneous dimerized couplings along the chain. This generalization preserves the model’s tractability while significantly enriching its structure. In particular, we derived closed-form expressions for the full energy spectrum and the corresponding eigenstates, providing tools that are essential for detailed theoretical analysis.

A natural next step is to investigate how the topological properties of the SSH model, along with other physically relevant quantities such as entanglement measures and correlation functions, are influenced by the introduction of these inhomogeneities. Since the doubling procedure provides a clear analytical handle on both the energy spectrum and the eigenstates, we expect that the computation and interpretation of these more intricate quantities will remain accessible through analytical methods. We will explore these aspects in detail in a forthcoming paper \cite{ssh_phys}.

Finally, it would be insightful to determine whether other physically relevant models can be analyzed using the approach proposed in this paper. For instance, the Kitaev chain with alternating bulk chemical potential, as explored in \cite{wouters2018exact}, could be generalized and solved exactly. Furthermore, the doubling of bivariate orthogonal polynomials—a topic not yet studied in detail—could facilitate the investigation of higher-dimensional inhomogeneous topological models. This would allow for a direct comparison with previous exact results for the $n$-dimensional SSH model \cite{Feng23}. Additionally, extending this study to the inhomogeneous SSH model with periodic boundary conditions would be of great interest; however, the theory of the required special functions is currently less developed and necessitates further mathematical research.

\vspace{0.5cm}

\noindent
\textbf{Acknowledgments: } N.~Cramp\'e is partially supported by the international research project AAPT of the CNRS. L.~Vinet is funded in part by a Discovery Grant from the Natural Sciences and Engineering Research Council (NSERC) of Canada. Q. Labriet and L. Morey enjoy postdoctoral fellowships provided by this grant.

\providecommand{\href}[2]{#2}\begingroup\raggedright\endgroup

\end{document}